# First-principles insights into the optoelectronic and thermoelectric properties of $X_3NbY_4$(X= Cu, Ag, Au; Y=S, Se, Te) sulvanite compounds for energy applications

Sadeya Sabnam Emo[1], Md. Sharear Aman[1], Md. Abdur Rashid[1*], and Jaker Hossain[2*]

[1]Department of Physics, University of Rajshahi, Rajshahi 6205, Bangladesh.
[2]Photonics & Advanced Materials Laboratory, Department of Electrical and Electronic Engineering, University of Rajshahi, Rajshahi 6205, Bangladesh.

## Abstract

The structural, electronic, optical and transport properties of $X_3NbY_4$(X= Cu, Ag, Au; Y=S, Se, Te) sulvanite chalcogenides materials have been investigated using the Full Potential Linear Augmented Plane wave (FP-LAPW) within the density functional theory (DFT). The calculated structural information of $X_3NbY_4$ compounds is consistent with reported results of the same family compounds. The electronic band diagram exhibit indirect type band structures with bandgap value in the range of $E_g$ ~1.65- 0.50 eV using PBE-GGA functional and 1.80 eV-1.18 eV using TB-mBJ functional which indicates that these are semiconductor materials. The density of states (DOS) shows that the amount of bandgap decreases owing to move of valence band maximum (VBM) to the high energy level whereas the conduction band minimum (CBM) to the low energy level owing to the replacement of $S \rightarrow Se \rightarrow Te$ and $Cu \rightarrow Ag \rightarrow Au$ atoms. The hybridized orbital by X-$d$, Nb-$d$ and Y-$p$ atomic orbitals dominate the VBM while hybridized by Nb-$d$ and Y-$p$ atomic orbitals mainly contribute the CBM. The elastic calculations exhibit that Cu-based materials have brittleness nature whereas Ag- and Au-based compounds are ductile nature. Furthermore, the phonon dispersion curves probes that these $X_3NbY_4$ compounds are dynamically stable. However, the calculated optical properties: dielectric function, absorption coefficient, refractive index, and energy loss function; specifically, the higher value of absorption coefficient (~$10^5$ cm$^{-1}$) indicates that these materials are attractive candidates in optoelectronics applications. Finally, thermoelectric parameters such as Seebeck coefficient, thermal conductivity, electrical conductivity, power factor (P.F) and ZT value of these compounds have also been investigated.

Overall, the finding explores that these materials are potential candidates for the applications in optoelectronic and thermoelectric devices.

## 1. Introduction

Owing to rising the worldwide energy demand as well as energy crises for lasting renewable energy resources has accelerated intensive research on novel semiconductor materials for the applications with a wide range in various industries such as photovoltaic (PV), solar cells, as well as thermoelectric (TE) devices. Among the various semiconductor materials, a large number of Cu/Ag/Au-based ternary chalcogenide materials have come more potential alternative materials of photovoltaic and thermoelectric devices applications broadly [1-7]. Recently, the chalcogenides semiconductors have fascinated important applications owing to its distinct band structures, optical absorption in a wide photon energy range, carrier transport properties, earth-abundance, non-toxicity, and tunable optoelectronic properties, making them potential alternative candidates for the high-efficiency thin-film PV applications. The Si-based materials has long been cornerstone for the PV applications with single-crystal, which have reached 25.6% efficiency due to the conversion from incident light energy to electric energy [8]. Owing to the bandgap of 1.1 eV of Si, its extreme efficiency is 30% as the Shockley–Queisser limit [9]. Due to the limit of Si-based devices, investigate on PV resources has expanded to progress other materials appropriate to solar cell applications.

Sulvanite type materials of $Cu_3MX_4$(where M = V, Nb, Ta and X = S, Se, Te) have gained attention as promising materials for PV, ion conductors, transparent conductors, and photocatalysts [2-4] and due to their favorable optoelectronic properties, including relatively wide optical bandgaps, inherent p-type conductivity, and photoemission in the visible light range [10,11]. First described by Pauling in 1933, the sulvanite crystal structures consist of interconnected $CuX_4$ and $MX_4$ tetrahedra, forming a robust and versatile framework [12,13]. Significant improvement has been made in the development of solution-processed chalcogenide materials for optoelectronic applications, offering a low-cost and scalable pathway for devices such as solar cells [14].

According to the literature, most studies have been concentrated on Cu-based chalcogenide compounds. Among these, various experimental groups have characterized on Cu-based sulvanite compounds such as $Cu_3VS_4$, $Cu_3NbS_4$, $Cu_3VSe_4$, $Cu_3NbSe_4$, $Cu_3TaSe_4$, $Cu_3NbTe_4$, and $Cu_3TaTe_4$ materials using numerous approaches including X-ray diffraction data via Rietveld method, Scanning Electron Microscopy (SEM), Energy-Dispersive X-ray spectroscopy (EDX), UV using spectrophotometry [15–20]. Simultaneously, several studies have been investigated on their structural information, thermodynamic, electronic and optical properties [3, ,5, 6, 21, 22, 23]. The experimental work has also been examined their mixed ionic-electronic conductivity behavior, as demonstrated in $Cu_3VS_4$ [24], along with vibronic properties probed by infrared reflectivity and Raman spectroscopy [25]. Additionally, thin-film transport and optical characteristics are investigated in detail of $Cu_3TaS_4$ [10,11].

On the other hand, to the best of our knowledge, theoretical or experimental studies on Ag/Au-based with Nb compounds adopting the space group $P\bar{4}3m$ of cubic structure remain are limited. The limitation motivates the investigation into the structural, elastic, electronic, optical as well as TE characteristics for Ag-/Au-based $I_3-V-VI_4$ semiconducting crystals in sulvanite type structure. The family of compounds $X_3NbY_4$(X= Cu, Ag, Au; Y=S, Se, Te) represents the promising class of materials for PV and optoelectronic applications. These materials are characterized by relatively wide optical bandgaps suitable for effective light absorption, intrinsic p-type conductivity facilitating efficient charge transport, and notable photoemission within the visible spectrum. Their unique combination of electronic and optical properties makes a position as a strong candidate for next-generation solar energy conversion and advanced optoelectronic devices applications. However, the Cu, Ag and Au-based sulvanite materials have become more attractive candidates in technological application area, but these are not explored broadly. To estimate silver and gold-based compounds to its coper-based materials, and measure their importance of energy harvesting applications. In this article, our goal is to investigate and compare these materials with their cooper-based analogues and measure their potential significance of PV devices applications.

In this work, a theoretical study has been investigated of $X_3NbY_4$ such as $Cu_3NbS_4$(CNS), $Cu_3NbSe_4$ (CNSe), $Cu_3NbTe_4$ (CNT), $Ag_3NbS_4$ (ANS), $Ag_3NbSe_4$ (ANSe), and $Ag_3NbTe_4$ (ANT), $Au_3NbS_4$ (AuNS), $Au_3NbSe_4$ (AuNSe) and $Au_3NbTe_4$ (ANT) via first-principles investigations on the DFT [26], implemented in Wien2k package [27,28]. The mechanical stability was assessed by estimating the elastic constants ($c_{ij}$), Young's modulus (Y), bulk modulus (B), shear modulus (G), Poisson's ratio (v) and Pugh's ratio (B/G). Due to the influence of elemental substitution Cu → Ag → Au, the electronic properties, photon energy dependence of various optical characteristic; dielectric function, absorption coefficient, refractive index, and optical conductivity across a broad range of incident photon energy have also been investigated. Moreover, to understand the transport behaviors including the Seebeck coefficient, electronic as well as lattice thermal conductivity, electrical conductivity, power factor and dimensionless quantity namely figures of merit ZT value are estimated via BoltzTraP scheme [29], from the Boltzmann kinetic transfer mechanism and rigid band approximation from the calculated band energy based on DFT. In these temperature dependent TE parameters calculations, a constant relaxation time is considered. Our findings divulge a middling energy gap value as well as distinct optical properties which make them potential materials could play the alternative with cooper-based materials for the future technology in PV, solar cells as well as in optoelectronic applications.

## 2. Computational methodology

To determine the physical structures and investigate the various properties, all investigation have been accomplished on FP-LAPW basis. In this research, DFT investigation have been performed via Wien2k code [27,28] to examine the structural, electronic, and optical characteristics of these compounds and the BoltzTraP code [29] was used to compute the TE characteristics. The structural characteristics have been estimated based on the Perdew–Burke–Ernzerhof generalized gradient approximation (PBE-GGA) for the exchange–correlation potential [30,31]. In contrast, the electronic properties and optical characteristics are evaluated via modified Becke–Johnson (mBJ) potential [27]. With FP-LAPW approach, maximum partial wave for the wavefunctions with a angular momentum quantum ($l_{max}$ = 10) are presented with spherical harmonics inside non-overlapping muffin-tin (MT) spheres enclosing the atomic sites. The space between the MT spheres is used as the basis set of plane wave for expanding the wavefunctions in a unit cell. For the plane wave expansion, cutoff value of $K_{max}$ = 7.0/RMT is used, where $K_{max}$ is the largest K-vector and RMT means the MT radius of used spheres. For the Fourier expansion, $G_{max}$ is specified as $12R_y^{1/2}$ for the charge density expansion. The Brillouin zone integrations have been computed using the Monkhorst–Pack package [32] the dense *k*-point (21×21×21) mesh of 10,000 points. The self-consistent field (SCF) computations are iterated by optimizing by the basics set (plane-wave) and a number of *k*-points until the energy converged to within $10^{-4}$ Ry. Firstly, the band structures, DOS calculations, and optical behaviors have been observed. The elastic parameters at ambient pressure have been estimated using Charpin tool via Wien2k code. In this approach, a small controlled strain has been applied in the cubic crystal. The resulting stress responses have been used to estimate the elastic constant $C_{ij}$ with the framework of DFT. The phonon band structure calculations have been computed via PHONOPY scheme [33] using DFT, employing a 2×2×2 supercell and a $\Gamma$-centered k-point mesh to measure the mechanical and dynamical stability of these sulvanite-structured compounds. This procedure ensured reliable convergence of the total energy. The transport behaviors have been assessed based on Boltzmann theory via BoltzTraP package [29], using the calculated band structure.

## 3. Results and discussions

### 3.1 Structural properties

To investigate the structural, electronic, dynamical, mechanical, and optical properties of transition metals of sulvanite structure chalcogenides compounds, a common crystal structure belongs space group $P\bar{4}3m$ (No. 215) with 8-atoms was adopted, as shown in Fig. 1.

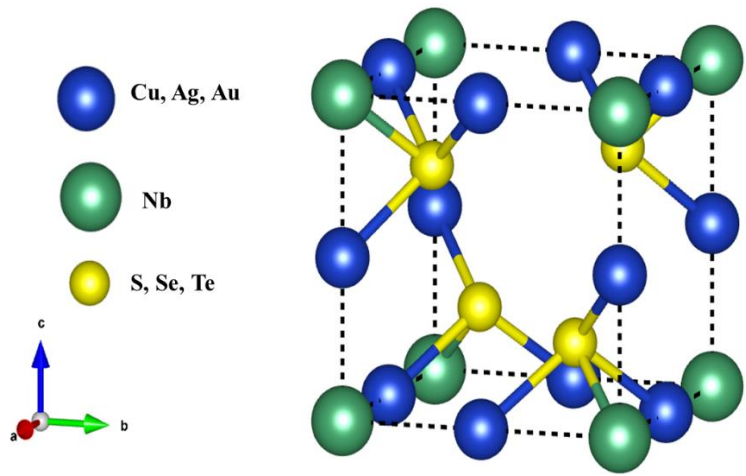

**Fig. 1:** Show the optimized structure of Cu-, Ag- and Au based of $X_3NbY_4$(X= Cu, Ag, Au; Y=S, Se, Te) cubic compounds. The Nb ions, X(Cu/Ag/Au) ions, and Y(S/Se/Te) elements are shown by green, blue and yellow color, respectively.

The initial structural parameters are adopted from Refs. [3,5-7,16-18]. Moreover, the unit cell parameters of $X_3NbY_4$(X= Cu, Ag, Au; Y=S, Se, Te) compounds has been further optimized. The optimized lattice constant and volume are tabulated in Table 1 and these parameters had been used to compute the required properties. The atomic positions correspond to Wyckoff sites: of transition metals X(Cu, Ag, Au) at 3d (0.5, 0.0, 0.0), Nb at 1a (0, 0, 0), and Y(S, Se, Te) at 4e (u, u, u), connected with point groups $D_{2d}$, $T_d$, and $C_{3v}$, respectively [6,18,19,23].

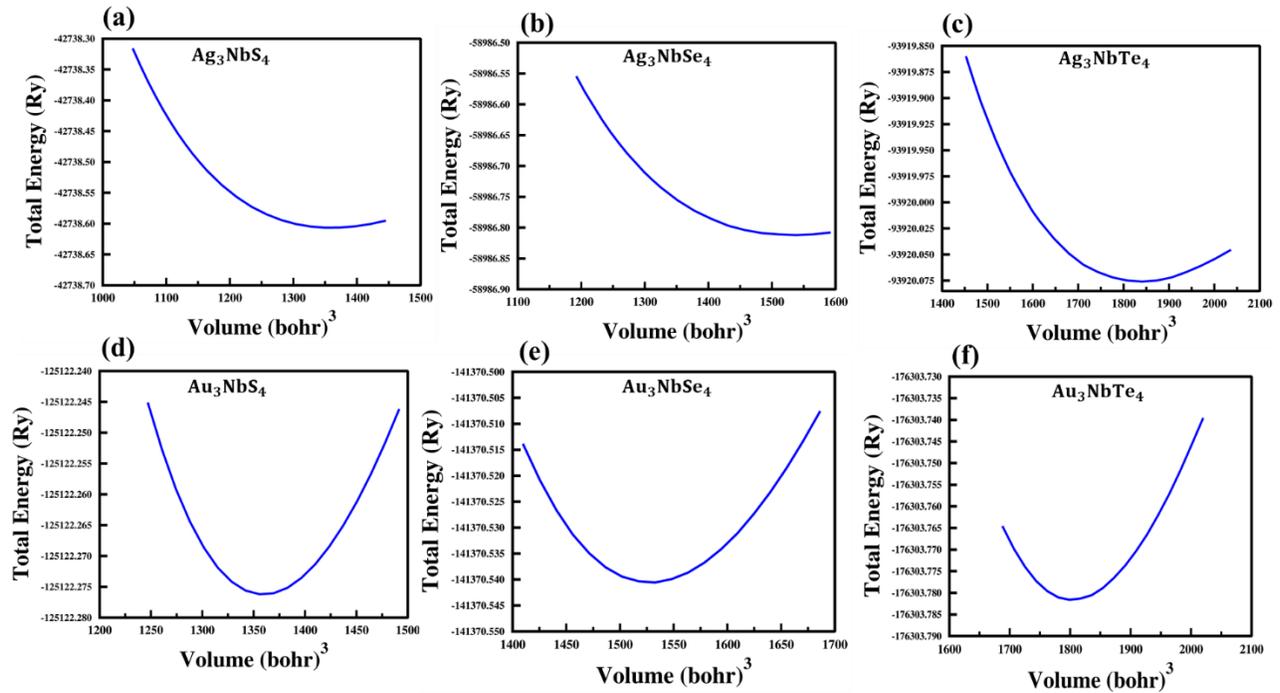

**Fig. 2:** (a-f) show the total energy in Ry vs volume in bohr$^3$ curves of $Ag_3NbS_4$, $Ag_3NbSe_4$, $Ag_3NbTe_4$, $Au_3NbS_4$, $Au_3NbSe_4$ and $Au_3NbTe_4$ chalcogenides compounds, respectively.

The fractional coordinates of (0.2420, 0.2420, 0.2420) for all compounds have been used. The lattice constants rise with atomic radius of the transition atoms (X) as well as the chalcogen elements (Y), following the sequence: a(CNS)<a(CNSe) < a(CNT)<a(ANS)< a(ANSe) < a(ANT)< a(AuNS) < a(AuNSe) < a(AuNT). This trend reflects the expected expansion of the unit cell owing to larger atomic radii, consistent with previous reports [3,6,15-19]. The volume with energy curves of $X_3NbY_4$(X= Cu, Ag, Au; Y=S, Se, Te) specimens have been depicted in the Fig.2 but Cu-based compounds are not presented because the cooper-based materials are examined extensively [16-18, 20,21,34,35].

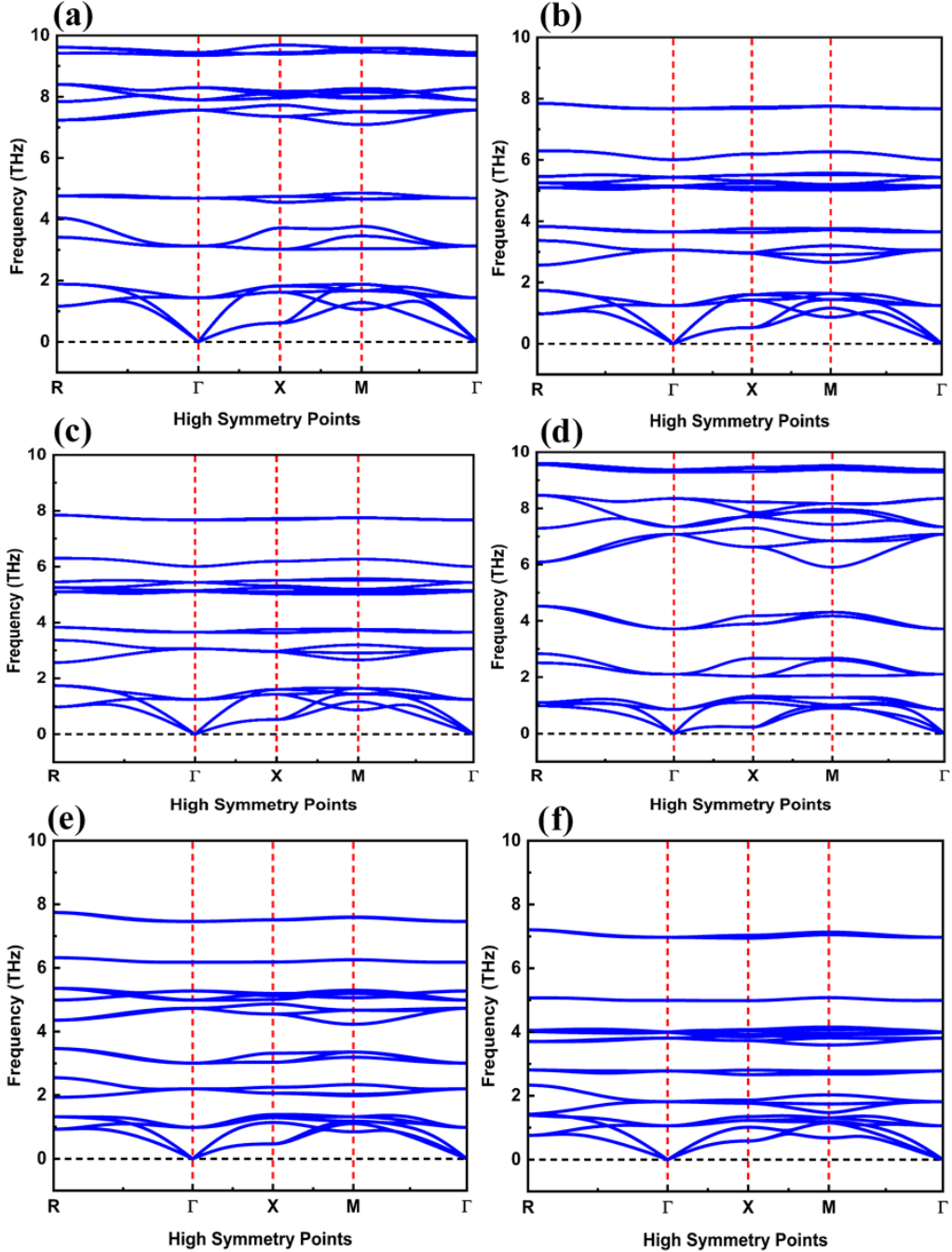

**Fig. 3:** (a-f) show the phonon dispersion curves (PDCs) with frequency in THz of $Ag_3NbS_4$, $Ag_3NbSe_4$, $Ag_3NbTe_4$, $Au_3NbS_4$, $Au_3NbSe_4$ and $Au_3NbTe_4$ chalcogenides compounds, respectively.

**Table 1:** The optimized lattice parameters a in Å, volume V Å$^3$ and energy in Ry of A$_3$TMB$_4$ (A=Cu, Ag, Au, TM=V, Nb, Ta, and B=S, Se, Te) sulvanite compounds.

| Compounds | $a$ (Å) | $V$ (Å$^3$) | Total energy (Ry) | Refs. |
|---|---|---|---|---|
| Cu$_3$NbS$_4$ | 5.5672 | 172.60 | -20731.23 | This |
| Cu$_3$NbSe$_4$ | 5.8129 | 196.43 | -36963.64 | This |
| Cu$_3$NbTe$_4$ | 6.2083 | 239.30 | -71877.25 | This |
| Cu$_3$NbS$_4$ | 5.5001, 5.472 | 166.38, ---- | --- | [16,3] |
| Cu$_3$NbSe$_4$ | 5.638, 5.641 | 179.22, ---- | --- | [17,3] |
| Cu$_3$NbTe$_4$ | 5.902, 5.9217 | ----, 205.65 | --- | [18, 3] |
| Cu$_3$VS$_4$ | 5.33 | 150.90 | -777.024 | [6] |
| Cu$_3$VSe$_4$ | 5.48 | 164.31 | -1546.74 | [6] |
| Cu$_3$VTe$_4$ | 5.804 | 195.57 | -626.79 | [6] |
| Ag$_3$NbS$_4$ | 5.8568 | 201.67 | -42690.77 | This |
| Ag$_3$NbSe$_4$ | 6.1012 | 227.07 | -58923.23 | This |
| Ag$_3$NbTe$_4$ | 6.4794 | 272.04 | -93836.91 | This |
| Ag$_3$VS$_4$ | 5.514 | 198.43 | -1102.83 | [6] |
| Ag$_3$VSe$_4$ | 6.002 | 216.26 | -1179.96 | [6] |
| Ag$_3$VTe$_4$ | 6.268 | 246.32 | -1112.34 | [6] |
| Ag$_3$TaS$_4$ | 5.904631 | 205.8630 | ------ | [5] |
| Ag$_3$TaSe$_4$ | 6.060895 | 221.54275 | ------ | [5] |
| Ag$_3$TaTe$_4$ | 6.287376 | 248.54691 | ----- | [5] |
| Au$_3$NbS$_4$ | 5.8622 | 201.48 | -125040.50 | This |
| Au$_3$NbSe$_4$ | 6.0954 | 226.49 | -141272.95 | This |
| Au$_3$NbTe$_4$ | 6.4411 | 267.23 | -176186.61 | This |
| Au$_3$VS$_4$ | 5.655 | 180.84 | ------- | [7] |
| Au$_3$VSe$_4$ | 5.986 | 214.49 | ------- | [7] |
| Au$_3$VTe$_4$ | 6.439 | 266.965 | ------- | [7] |

The estimated formation energy using $E_{for} = E_{X_3NbY_4} - (3E_X + E_{Nb} + 4E_Y)$ are found -41.429, -36.776, -31.728, -29.361, -25.116, -20.640, -9.538, -5.782 and -1.129 eV/unit cell of the CNS, CNSe, CNT, ANS, ANSe, ANT, AuNS, AuNSe, and AuNT materials, respectively. The negative value of $E_{for}$ show that the creation of the considered specimens is dynamically stable. Moreover, the phase of $X_3NbY_4$(X= Cu, Ag, Au; Y=S, Se, Te) cubic compounds is chemically and dynamically stable. The calculated phonon dispersion curves (PDC) are plotted in Fig. 3(a–f) of ANS, ANSe, ANT, AuNS, AuNSe, and AuNT semiconducting materials, respectively. According to the Fig. 3(a-f), negative phonon frequency is not found of these curves of $X_3NbY_4$ compounds which confirm their dynamical stability [36-38]. The unit cell of $X_3NbY_4$ contains 8 atoms; the 24 phonon modes are appeared that leads the 21 and 3 optical and acoustic modes respectively. Therefore, these $X_3NbY_4$(X= Cu, Ag, Au; Y=S, Se, Te) semiconducting materials are stable in dynamically as well as chemically [38,39].

## 3.2 Elastic properties

The elastic stiffness constants and their related different parameters deliver information of the mechanical behavior, stability and stiffness of the materials. The recorded three independent elastic parameters such as $C_{11}$, $C_{12}$, and $C_{44}$ in a cubic crystal in the form $X_3NbY_4$ semiconductors are listed in the Table 2 which are good consistent with reported results [5,6,19,40]. The recorded elastic constants satisfy the stability proviso according to Born–Huang approach [41]. The following situations are satisfied: $C_{11} - C_{12} > 0$, $C_{11} + 2C_{12} > 0$ and $C_{44} > 0$, cubic crystals are mechanically stable. Cauchy pressure, $C_{cauchy} = (C_{12} - C_{44})$ is estimated which measure the ductility or brittleness [6,42,43]. According to the estimated value of $C_{cauchy}$, the Cu-based materials exhibit the brittleness nature whereas Ag and Au-based compounds show the ductile nature. The recorded value of machinability index ($\mu^M$), B, G, Y, B/G and v [44-46] are tabulated in the Table 2. The numerical value for $B$, G, E, and v have been measured by using well-known the Reuss, Voigt and Hill theories [47-49]. According to the Table 2, the value of Y decreases with the Cu → Ag → Au as well as S → Se → Te atom replacement in the $X_3NbY_4$(X= Cu, Ag, Au; Y=S, Se, Te) semiconductors. According to table 2, the value of B/G of Cu-based materials shows less than 1.75 which indicates the brittleness whereas Ag and Au-based materials exhibit greater than 1.75 which indicate the ductile behavior [50-52]. Consequently, the Ag and Au-based chalcogens are appropriate candidates in applications for demanding as flexible solar cells whereas Cu-based samples are not suitable for flexible energy harvesting applications. The value of Poisson's ratio which is another indicator for measuring the ductile or brittleness characteristic belongs in range from 0.25 to 0.35, the compounds display ductile behavior; otherwise, brittleness. As shown in table, the Cu-based materials also shows brittle nature whereas Ag and Au-based materials display the ductile behavior. The machinability index, $\mu^M(B/C_{44})$ mirrors the material's flexible deformability and dry lubricating nature. The calculated value of $\mu^M$ recommend that these materials have promising machinability features.

**Table 2:** Calculated value of elastic parameters $C_{ij}$, Cauchy pressure $C_{cauchy}$, bulk modulus $B$, shear modulus $G$, Young's modulus Y in GPa, Pugh's ratio $B/G$, Poisson's ratio $v$, machinability index $\mu^M$ of $X_3NbY_4$(X= Cu, Ag, Au; Y=S, Se, Te) semiconductors.

| Compounds | $C_{11}$ | $C_{12}$ | $C_{44}$ | $C_{cauchy}$ | B | G | Y | B/G | $v$ | $\mu^M$ |
|---|---|---|---|---|---|---|---|---|---|---|
| CNS | 139.44 | 59.69 | 58.33 | 1.36 | 86.25 | 50.2 | 125.91 | 1.72 | 0.24 | 1.47 |
| CNSe | 90.64 | 47.51 | 51.16 | -3.65 | 27.89 | 21.32 | 50.97 | 1.30 | 0.19 | 0.54 |
| CNT | 81.78 | 46.96 | 49.80 | -2.84 | 58.57 | 33.70 | 82.20 | 1.73 | 0.24 | 1.17 |
| ANS | 117.8 | 52.97 | 48.13 | 4.84 | 74.61 | 41.10 | 104.45 | 1.81 | 0.25 | 1.55 |
| ANSe | 96.09 | 49.64 | 45.88 | 3.76 | 51.24 | 34.91 | 88.98 | 1.86 | 0.27 | 1.12 |
| ANT | 70.71 | 41.48 | 40.87 | 0.61 | 82.93 | 27.47 | 69.88 | 1.856 | 0.27 | 2.03 |
| AuNS | 129.07 | 59.86 | 47.71 | 12.15 | 72.65 | 41.95 | 107.7 | 1.97 | 0.28 | 1.52 |
| AuNSe | 105.56 | 56.10 | 48.01 | 8.09 | 60.23 | 36.79 | 94.5 | 1.97 | 0.28 | 1.25 |
| AuNT | 81.92 | 59.48 | 46.75 | 12.73 | 59.23 | 30.68 | 78.88 | 1.99 | 0.28 | 1.27 |

### 3.3 Electronic Properties

The evaluated band structure in the first Brillouin zone (FBZ) of the CNS, CNSe, CNT, ANS, ANSe, ANT, AuNS, AuNSe, and AuNT semiconducting materials are depicted in Fig. 4(a-i), respectively. In each figure the horizontal line (dashed line in black colour) displays the Fermi level ($E_F$). The filled valence bands (VB) and the unfilled conduction bands (CB) are separated by an energy gap namely bandgap ($E_g$) which is the separation between the VB edge and CB edge.

**Table 3:** Bandgap value $E_g$ (eV) between the VBM and CBM of $A_3TMB_4$ (A=Cu, Ag, Au, TM=V, Nb, Ta, and B=S, Se, Te) compounds chalcogenides.

| Compounds | $E_g$ (eV) | | |
|---|---|---|---|
| | PBE-GGA | TB-mBJ | |
| $Cu_3NbS_4$ | 1.65 | 1.80 | This work |
| $Cu_3NbSe_4$ | 1.42 | 1.57 | |
| $Cu_3NbTe_4$ | 1.10 | 1.24 | |
| $Cu_3VS_4$, $Cu_3NbS_4$, $Cu_3TaS_4$ $Cu_3VSe_4$, $Cu_3NbSe_4$, $Cu_3TaSe_4$ $Cu_3VTe_4$, $Cu_3NbTe_4$, $Cu_3TaTe_4$ | 1.13, 1.30 [3,25], 1.82, 1.66 [3,23], 2.10, 2.70 [3,11] 0.87 [3], 1.45 [3], 1.71, 2.35 [3,11] 0.53 [3], 0.92 [3], 1.11 [3] | | |
| $Ag_3NbS_4$ | 1.60 | 1.75 | This work |
| $Ag_3NbSe_4$ | 1.36 | 1.56 | |
| $Ag_3NbTe_4$ | 1.05 | 1.23 | |
| $Ag_3TaS_4$, $Ag_3TaSe_4$, $Ag_3TaTe_4$ | 1.70, 150, 1.24 | ----- | [5] |
| $Ag_3VS_4$, $Ag_3VSe_4$, $Ag_3VTe_4$ | 0.86, 0.63, 0.47 | ----- | [6] |
| $Au_3NbS_4$ | 0.69 | 1.53 | This work |
| $Au_3NbSe_4$ | 0.51 | 1.25 | |
| $Au_3NbTe_4$ | 0.50 | 1.18 | |
| $Au_3VS_4$, $Au_3VSe_4$, $Au_3VTe_4$ | 0.84, 0.66, 0.45 | 1.00, 0.89, 0.55 | [7] |

The calculated amount of this $E_g$ indicates the indirect bandgap type semiconductors. The VB edge is attributed at the *R*-point for all specimen and the CB edge is appeared at X-point except AuNS and AuNSe compounds (*Γ*-point). The measured bandgap using PBE-GGA and TB-mBJ functional are tabulated in the Table 3. Actually GGA-PBE functionals decreases the value of bandgap between the VBM and CBM owing to the interactions for the valence electrons cloud and the core ions in crystals. But in this study, the calculated band structures via TB-mBJ have been shown because optical and thermoelectric properties are calculated using the TB-mBJ functionals.

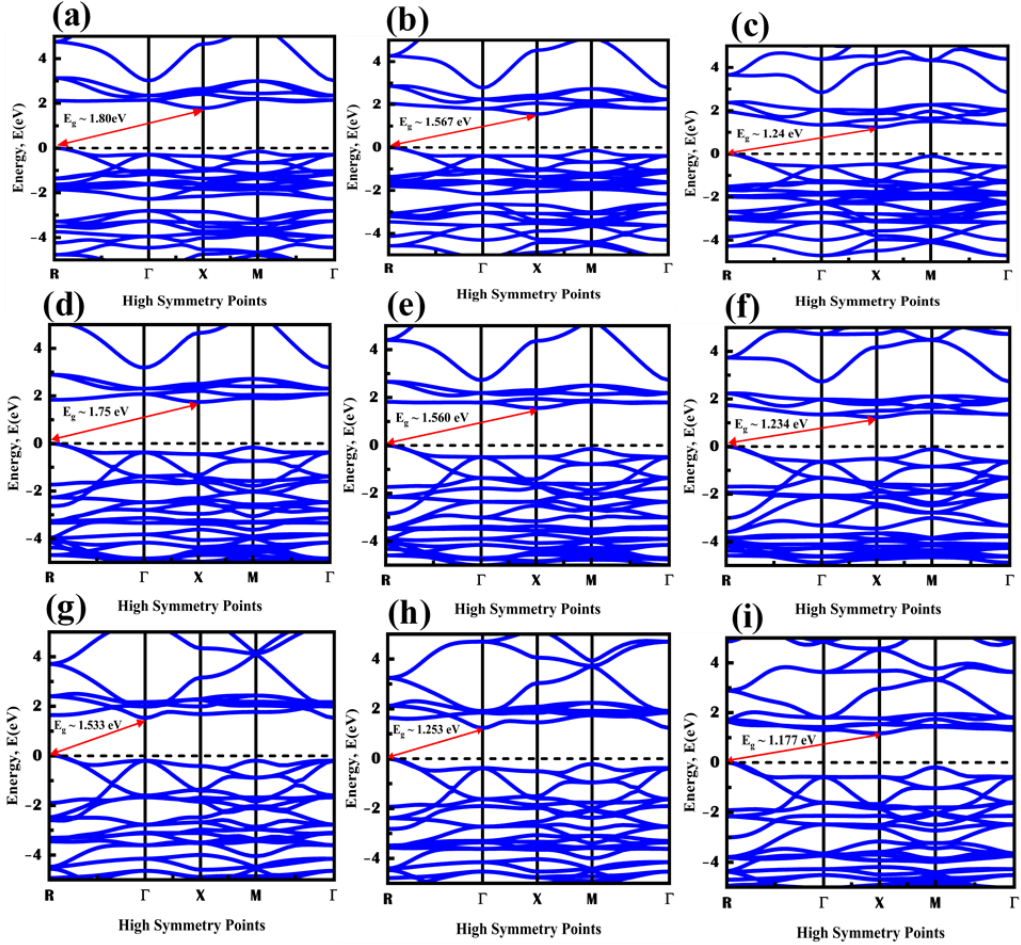

**Fig 4:** (a-i) show the electronic band structure of $Cu_3NbS_4$, $Cu_3NbSe_4$, $Cu_3NbTe_4$, $Ag_3NbS_4$, $Ag_3NbSe_4$, $Ag_3NbTe_4$, $Au_3NbS_4$, $Au_3NbSe_4$ and $Au_3NbTe_4$ chalcogenides compounds, respectively using TB-mBJ functional.

The $E_F$ stands at 0 eV energy level as a reference level that coincides at the edge of the VB. It is evident that the VBM and the CBM do not coincide with same momentum which exhibit the indirect bandgap with a bandgap value of 1.80, 1.57, 1.24, 1.75, 1.56, 1.23, 1.53, 1.25, 1.18 eV of CNS, CNSe, CNT, ANS, ANSe, ANT, AuNS, AuNSe, and AuNT semiconducting materials, respectively. According to band structure calculations, the numerical value of $E_g$ decreases with chalcogens element S→ Se →Te for fixed Cu or Ag or Au atom and also decreases for shifting the atom Cu→ Ag→Au with fixed chalcogen S or Se or Te atom that are consistent with the reported results [3,5,6]. The VB are significantly influenced by Cu/Ag/Au *d*- and S/Se/Te *p*-orbitals while CB are affected by Nb *d*- and S/Se/Te *p*-states in Cu/Ag/Au based chalcogenides, respectively. To elucidate the participation of different atomic orbitals, the total density of states (TDOS) and partial density of states (PDOS) have been analyzed. The computed TDOS and PDOS of CNS, CNSe, CNT, ANS, ANSe, ANT, AuNS, AuNSe, and AuNT materials are depicted in the Fig. 5(a-i), correspondingly.

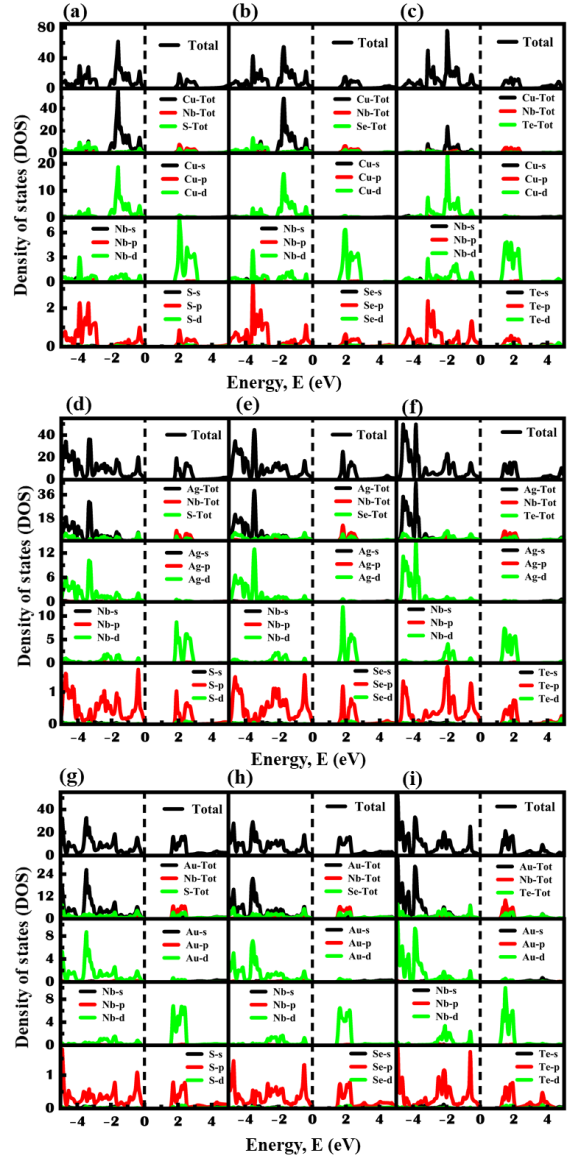

**Fig. 5:** (a-i) show the density of states (DOS), and partial density of states (PDOS) of $Cu_3NbS_4$, $Cu_3NbSe_4$, $Cu_3NbTe_4$, $Ag_3NbS_4$, $Ag_3NbSe_4$, $Ag_3NbTe_4$, $Au_3NbS_4$, $Au_3NbS_4$ and $Au_3NbTe_4$ compounds, respectively using TB-mBJ functional.

According to the Fig. 5(a-i), VB edge is originated by the hybridizing of Cu-*d*, Nb-*d* and S-*p* states and dominated by Cu-*d* orbital whereas CB edge is attributed by hybridization of Nb-*d* and S-*p* states and dominated by Nb-*d* orbitals for CNS, CNSe, CNT; VBM edge is originated by the hybridizing of Ag-*d*, Nb-*d* and S-*p* states and dominated by the S-*p* orbital whereas CBM is attributed by hybridizing of Nb-*d* and S-*p* states and dominated by Nb-*d* states for ANS, ANSe, ANT; VBM is originated by the hybridizing of Au-*d*, Nb-*d* and S-*p* states and dominated by the Au-*d* states whereas CBM is attributed by hybridization of Nb-*d* and S-*p* states and dominated by Nb-*d* orbitals for AuNS, AuNSe, and AuNT semiconducting materials.

## 3.4 Optical Properties

The various optical behaviors of any materials explain the interaction between the matter and incident electromagnetic radiation (EMR). This interaction of EMR with matter measure the various dimension of optoelectronic devices applications. In this section, the incident energy dependence various optical properties including the dielectric functions ($\varepsilon(\omega)$), absorption coefficient ($\alpha(\omega)$), refractive index ($\eta(\omega)$), extinction coefficient ($\kappa(\omega)$), optical conductivity ($\sigma_{opt}(\omega)$), reflectivity (R($\omega$)) and energy loss function (L($\omega$)) have been investigated. Ehrenreich and Cohen's equation, $\varepsilon(\omega) = \varepsilon_1(\omega) + i\varepsilon_2(\omega)$ is used to explain the interaction between photon energy and matter [53], where $\varepsilon_1(\omega)$ represents the real part which explain the electronic polarization and anomalous dispersion in the samples and $\varepsilon_2(\omega)$ also represents the imaginary part which describe the absorption in materials.

The calculated value of $\varepsilon_1(\omega)$ and $\varepsilon_2(\omega)$ of the $\varepsilon(\omega)$ with incident EMR in the range 0 to 12 eV have been shown in the Fig. 6(a-c). In Fig. 6(a-c), the energy dependence black, red and green curves for the CNS, CNSe and CNT compounds, respectively. The solid and dashed curves represent for the $\varepsilon_1(\omega)$ and $\varepsilon_2(\omega)$, respectively. The computed value of the static dielectric constants $\varepsilon_1(0)$ are recorded about 6.0, 7.4, 8.5, 6.2, 6.9, 8.0, 7.9, 8.5 and 9.0 of CNS, CNSe, CNT, ANS, ANSe, ANT, AuNS, AuNSe, and AuNT semiconductors, respectively which are good consistent with the reported results [5,6, 40, 54].

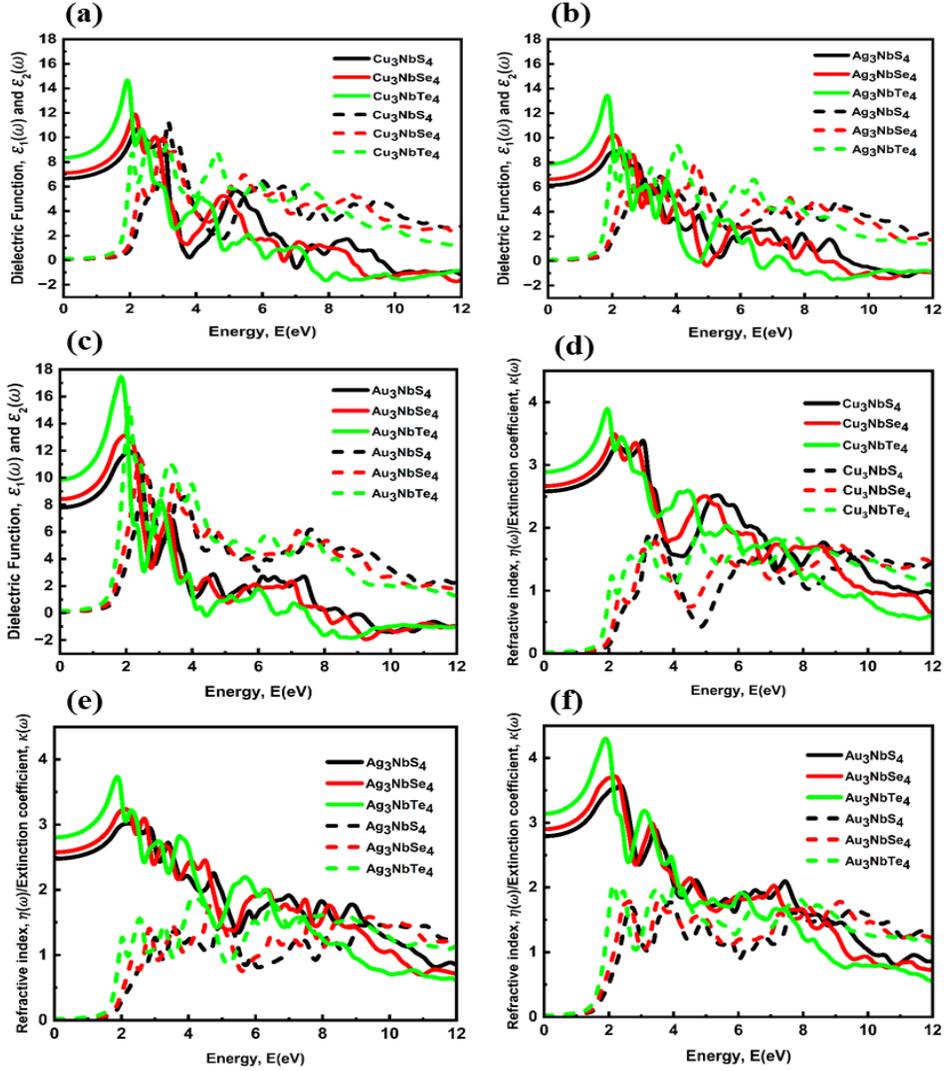

**Fig. 6:** Show the computed value of (a-c) real part $\varepsilon_1(\omega)$ of $\varepsilon(\omega)$ and imaginary part $\varepsilon_2(\omega)$ of $\varepsilon(\omega)$, (d-f) refractive index η(ω) and the extinction coefficient k(ω) in the range 0 to 12 eV incident light energy of $X_3NbY_4$ semiconductors.

The estimated value suggest that these semiconductors exhibit different dielectric characteristic. The peak value of $\varepsilon_1(\omega)$ are found in the visible range at 10.0 at 2.4 eV, 11.5 at 2.2 eV, 14.2 at 2.0 eV, 8.5 at 2.0 eV, 9.0 at 2.0 eV, 13.9 at 1.9 eV, 11.5 at 2.0 eV, 12.5 at 2.0 eV and 17.2 at 1.9 eV of CNS, CNSe, CNT, ANS, ANSe, ANT, AuNS, AuNSe, and AuNT semiconductors, respectively.

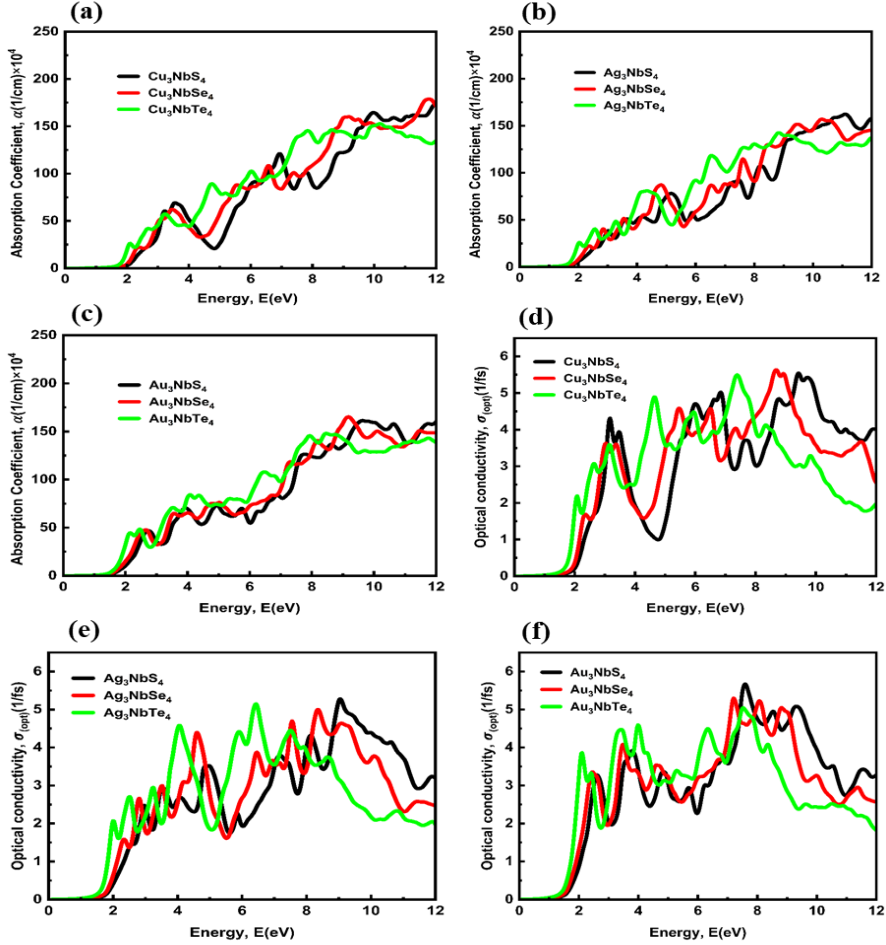

**Fig. 7:** Show the calculated value of (a-c) the absorption coefficients, $\alpha(\omega)$ and (d-f) conductivity $\sigma_{opt}(\omega)$ in the range 0 to 12 eV incident light energy of $X_3NbY_4$ semiconductors.

These results highlight these semiconductors as a potential material for the optoelectronic devices application in the visible light range. Fig. 6(d-f) shows the $\varepsilon_2(\omega)$ of the energy dependance dielectric function of cooper, silver and gold-based materials, correspondingly. The computed beginning (threshold) photonic energy of $\varepsilon_2(\omega)$ have been observed about 1.9, 1.6, 1.4, 1.8, 1.6, 1.3, 1.62, 1.32 and 1.22 eV of CNS, CNSe, CNT, ANS, ANSe, ANT, AuNS, AuNSe, and AuNT semiconductors, respectively. In the visible region (VR), $\varepsilon_2(\omega)$ exhibits several prominent peaks with incident photon energy that are corresponding to the interband optical transitions.

The recorded energy dependence refractive index η(ω) and the extinction coefficient k(ω) are demonstrated in the Fig. 6(d-f) Cu-, Ag- and Au-semiconductors. respectively. The η(ω) is a crucial characteristic of materials for the proposal of optoelectronics device including the waveguides, optical fibers and various lenses. The k(ω) is another vital property which governs the incident photon absorption for the photon energy conversion. According to the Fig. 6(d-f), the static value of η(0) are estimated about 2.5, 2.7, 2.9, 2.4, 2.6, 2.8, 2.7, 2.8 and 3.4 of CNS, CNSe, CNT, ANS, ANSe, ANT, AuNS, AuNSe, and AuNT semiconductors, respectively which follow the relation of the η (0) of the $\varepsilon_1(0)$ of static dielectric value, η² (0) ≈ $\varepsilon_1(0)$ [55,56]. The maximum amount of η(ω) have been obtained in VR which are similar with the reported results [57]. The value of η(ω) reductions with rises the EMR energy. These important properties make these compounds appropriate for the device's applications in this range. A greater k(ω) specifies stronger optical absorption in material. The difference of η(ω) thoroughly follows the trend with the $\varepsilon_1(\omega)$, whereas the value of k(ω) reflects the value of the $\varepsilon_2(\omega)$. Estimated absorption coefficients, $\alpha(\omega)$ with EMR energy in the range 0 to 12 eV have been showed in the Fig. 7 (a-c) for cooper, silver as well as gold containing semiconductors, respectively. Absorption edges namely optical bandgaps which are the pointer of the electronic bandgap $E_g$. The highest value of $\alpha(\omega)$ got approximately (4 to 7) ×$10^5$ cm$^{-1}$ in the incident light range (VR) for all semiconductors. Therefore, estimated results display the highest absorption which are good candidate for the solar application in the devices. The calculated the optical conductivity $\sigma_{opt}(\omega)$ with energy (photon energy) have been shown in the Fig. 7(d-f) for the cooper, silver and gold containing semiconductors, correspondingly. According to the Fig. 7(d-f), the amount of the $\sigma_{opt}(\omega)$ stay zero at the lowest energy which confirm the semiconducting nature. Its value increases corresponding to the value of bandgap and the highest value of it in low-energy range are detected. These results reveal of these Cu, Ag and Au-based semiconductors as a good material for the devices application in photovoltaic cells owing to their higher optical conductivity in VR and also in UV region.

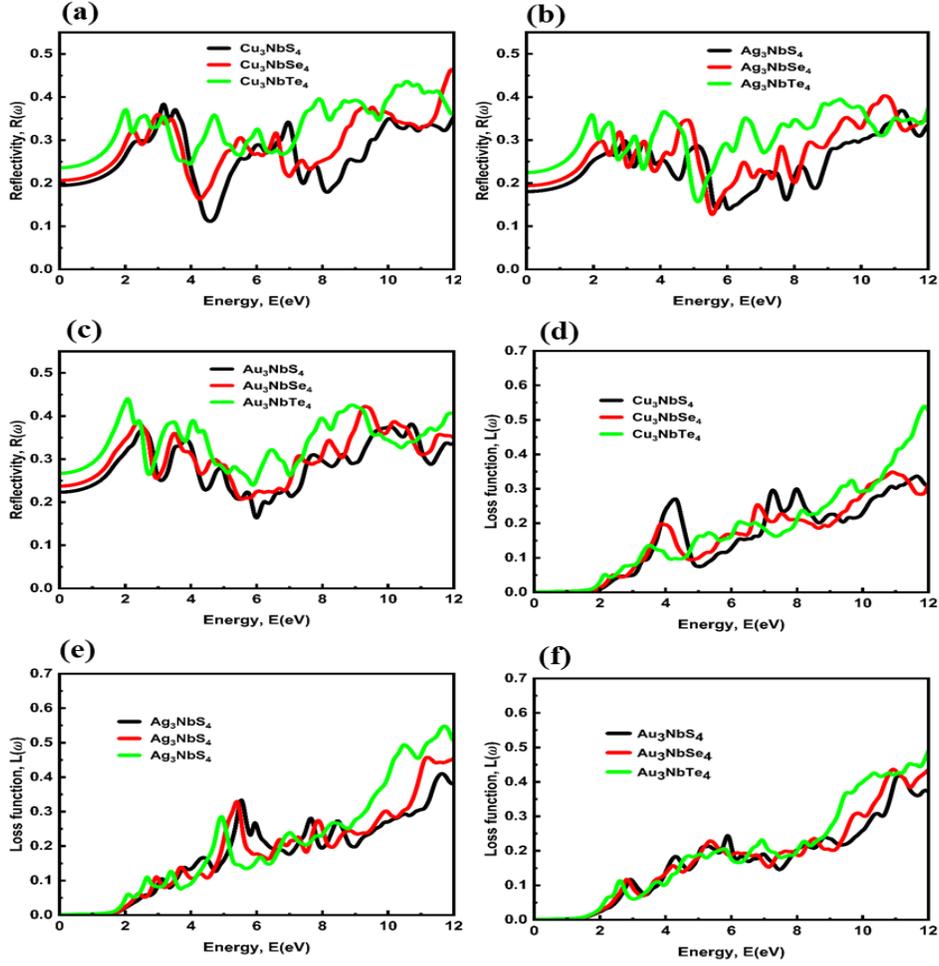

**Fig. 8:** Show the calculated value of (a-c) the variation of reflectivity $R(\omega)$ and (d-f) the variation of energy loss function $L(\omega)$ with energy of $X_3NbY_4$ semiconductors.

The parameters reflectivity $R(\omega)$ is the crucial for the coatings, mirrors and reflective technologies applications that pronounce the reflections of light (A fraction) from the materials surface. The estimated amount of $R(\omega)$ is exhibited in the Fig. 8(a-c) of cooper, silver and gold containing compounds, correspondingly. According to the 6(a-c), the relatively low value of $R(\omega)$ of these semiconductors which specify the transparent coating devices applications in the VR. The value of $L(\omega)$ of materials explain interactions of EMR with matter.

The calculated value of $L(\omega)$ have been display in the Fig. 8(d-f) of Cu, Ag as well as Au-based compounds, respectively. According to these figures, the value of $L(\omega)$ displays reasonably small photon energy loss in specific regions which permit them to remain the incident light energy integrity. As shown in the Fig. 8(d-f), the value of L($\omega$) are found approximately 0.1, 0.25 and 0.11 at 2.6 eV, 4.1 eV and 4.5 eV for CNS, 0.12, 0.2, 0.19 at 2.0 eV, 4.0 eV and 4.8 eV for CNSe, 0.1, 0.15 and 0.18 at 2.4 eV, 2.8 eV and 4.4 eV for CNT, 0.1, 0.12 and 0.30 at 2.2 eV, 4.2 eV and 4.6 eV for ANS, 0.11, 0.16 and 0.31 at 2.5 eV, 2.9 eV and 4.7 eV for ANSe, 0.12, 0.13 and 0.25 at 2.4 eV, 2.8 eV and 4.6 eV for ANT, 0.11, 0.16 and 0.21 at 2.5 eV, 4.3 eV and 5.8 eV for AuNS, 0.11, 0.20 and 0.25 at 2.5 eV, 4.2 eV and 6.0 eV for AuNSe and 0.12, 0.15 and 0.20 at 2.4 eV, 3.8 eV and 5.8 eV for AuNT semiconducting materials, respectively. These value of loss functions suggest that these Cu, Ag and Au-based semiconducting materials can be alternative potential candidates for optical materials [58].

## 3.5 Effective mass and Excitonic properties

The effective mass ($m^*$) of the carriers (electrons and holes) in semiconductors indicates how these carriers show the behaviors of mass owing to the crystal potential that is defined by the curvature of the corresponding band structure using this $m^* = \hbar^2 (d^2E/dk^2)^{-1}$ relation. The numerical value of effective mass has been estimated by *E-k* band structure curve fitting using the above-mentioned relation. The calculated reduced effective mass ($\mu$) with Rydberg energy ($E_b$) and Bohr radius ($r_B$) via this relation $\mu^{-1} = m_e^{*-1} + m_h^{*-1}$, where $m_e^*$ represents the effective mass of electrons and $m_h^*$ is the effective mass of holes. The computed value of effective mass of electrons at the *X*-point in the conduction band for all compounds, and holes at X-point for CNS, CNSe, CNT, ANS, ANSe, ANT, and AuNT whereas at $\Gamma$-point for AuNS, AuNSe compounds to corresponding valence band in the symmetry *k*-points have been estimated via the calculated band structures. Estimated value of this $m^*$ in table 4. The sufficient absorbed photon energy creates an exciton from the bound pair of electron-hole owing to the coulombic interaction in the materials. According to the Wannier–Mott exciton model, the value of exciton binding energy ($E_b$) have been computed from this relation, $E_b = R_\infty \frac{\mu}{m_0 \varepsilon_\infty^2}$, $R_\infty$ represent the Rydberg constant and its value equal to 13.6 eV, $\varepsilon_\infty$ means the static dielectric value, m$_0$ means mass (free electron). The exciton Bohr radius ($R_{ex}$) have also been computed using $R_{ex} = \frac{m_0}{\mu^*} \varepsilon_{eff} \, n^2 \, r_B$ formula, where *n* is the exciton level (energy), its value of 1 for lowest exciton radius, $r_B$ is the Bohr radius [6, 59]. The value of the excitonic binding energy is the important indicators in semiconducting materials for the application, naturally its value in the range few meV to 100 meV [6, 60]. The calculated value of exciton Bohr radius ($R_{ex}$) and exciton binding energy ($E_b$) of these compounds are shown in Table 4.

**Table 4:** Calculated effective mass $m^*$ in $m_0$ of electrons and hoes, reduce mass $\mu$ in $m_0$, Bohr radius $R_{ex}$ in Å and exciton binding energy $E_b$ in meV of X$_3$NbY$_4$ semiconductors.

| Compounds | $m_e^*$ in $m_0$ | $m_h^*$ in $m_0$ | $\mu$ in $m_0$ | $R_{ex}$ in Å | $E_b$ in meV |
|---|---|---|---|---|---|
| CNS | 0.541 | 1.093 | 0.3619 | 8.77 | 136.7 |
| CNSe | 0.591 | 0.769 | 0.3342 | 11.72 | 83.0 |
| CNT | 0.841 | 0.655 | 0.3682 | 12.21 | 69.3 |
| ANS | 0.893 | 0.884 | 0.4420 | 7.38 | 157.2 |
| ANSe | 1.44 | 1.114 | 0.6281 | 5.82 | 179.3 |
| ANT | 0.797 | 1.248 | 0.4864 | 8.70 | 103.3 |
| AuNS | 0.167 | 0.43 | 0.1203 | 34.74 | 26.2 |
| AuNSe | 0.205 | 0.312 | 0.1240 | 36.35 | 23.3 |
| AuNT | 0.568 | 0.229 | 0.1630 | 29.17 | 27.4 |

The lower exciton binding energy of compounds display the greater charge separation capacity and this material are the perfect for PV devices technology, on the other hand, higher exciton binding energy offer the light-emitting devices (LEDs) technology [6, 61]. As shown in Table 4, Au containing compounds are good materials for PV and optoelectronic devices and two materials are excellent for the LEDs, solar cells owing to the greater binding energy. Therefore, atomic substitutions of Cu → Ag→Au and S → Se → Te changes the excitonic behaviors which suggest these considered materials are good various the optoelectronic technologies.

## 3.6 Transport properties

Thermoelectric (TE) behaviors in thermoelectric material (TEM) describe the contributions of the structural and electronic nature in their efficiency in energy conversion of heat energy into electric energy due to temperature gradient. The efficiency of TEM is measured by dimensionless quantity (figure of merit) specifically ZT value. The numerical value of ZT is computed via $ZT = \left(\frac{S^2\sigma}{\kappa}\right)T$, where S, $\sigma$, T and $\kappa$ are the Seebeck coefficient, electrical conductivity, absolute temperature and total thermal conductivity. The value of $\kappa$ depends on the lattice thermal ($\kappa_L$) and electronic thermal ($\kappa_e$) conductivities. The Slack model, [62] was employed to estimate the value of $k_L(T) = A(\gamma)\frac{M_{av}\theta_D^3\delta}{T\gamma^2 n^{2/3}}$; $[A(\gamma) = \frac{5.720\times 10^7 \times 0.849}{2\left(1-\frac{0.514}{\gamma}+\frac{0.228}{\gamma^2}\right)}]$, where, $M_{av}$, $\theta_D$, $\delta$, T, $\gamma$ and n present the average atomic mass (kg/mol), Debye temperature (K), cubic root of average atomic volume (m), temperature (K), Grüneisen parameter and , number of atoms in the conventional unit cell, and the parameter $\gamma$ is measured using Poisson's ratio (ν), $\gamma = \frac{3(1+\nu)}{2(2-3\nu)}$ . Furthermore, a constant relaxation time approximation $\tau = 10^{-14}$ s [63,64] was assumed to assess the ZT value. The excellent TEMs belongs higher value of S and $\sigma$ but lower value of $\kappa$ for maintaining the temperature gradient. The computed various TE parameters of $X_3NbY_4$(X= Cu, Ag, Au; Y=S, Se, Te) with variation of temperature from 300 K to 950 K are presented in Fig. (9-11).

According to the Fig. 9(a-c), the estimated value of the Seebeck coefficient reduces with the variation of temperature of Cu- and Ag-based semiconductors up to 900 K which are good consistent with the reported results [5, 6, 65]. The measured maximum value of S approximately 0.259 mV/K (CNS), 0.245 mV/K (CNSe), 0.234 mV/K (ANS), 0.255 mV/K (ANSe) and 0.256 mV/K (ANT) at 300 K but 0.235 mV/K (CNT) at 400 K. On the other hand, the variation of S with temperature of Au-based compounds is negligible. It is also noted that the value of S for AuNT compound is very low. These results suggest that all considered compounds can be potential TEMs except AuNT compound because the mobility of electrons or holes depends on the value of Seebeck coefficient. The higher value of Seebeck coefficient in TEMs leads greater mobility.

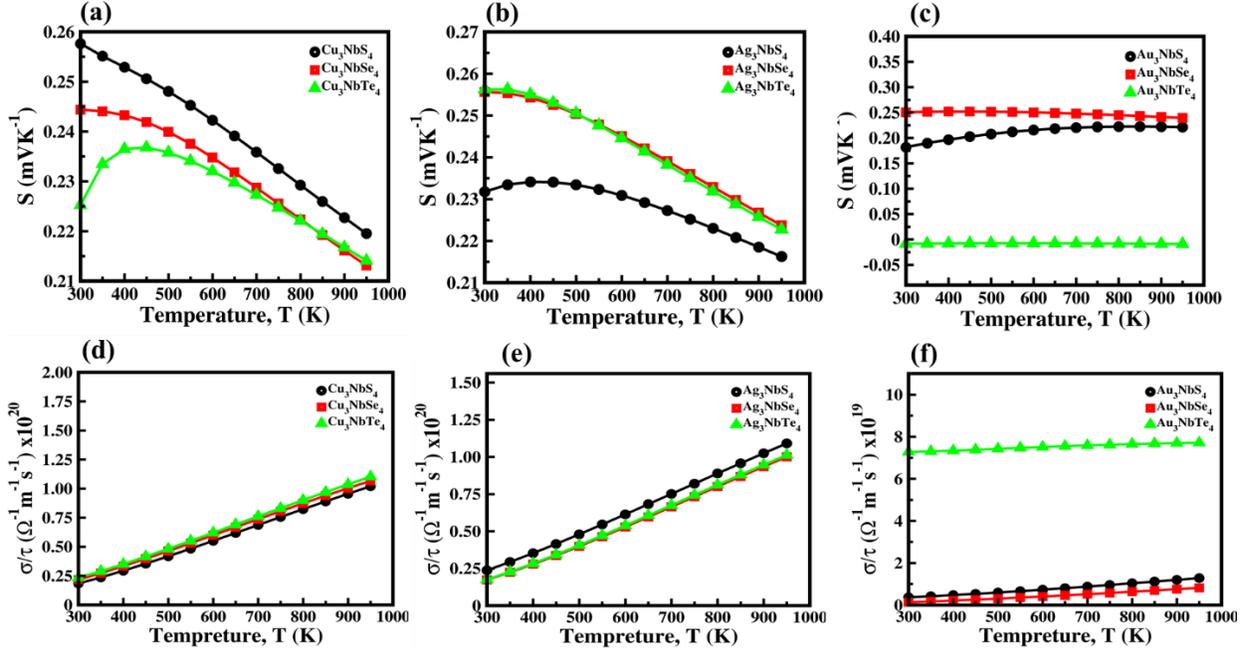

**Fig. 9:** Show the variation (a-c) of Seebeck coefficient S(T) and (d-f) of conductivity (electrical) $\sigma(T)$ up to 950 K temperature for Cu-, Ag- and Au- based compounds, respectively, where CNS, ANS, AuNS in black curves, CNSe, ANSe, AuNSe in red curves, CNT, ANT, AuNT in green curves, respectively.

Fig. 9(d-f) display the value of $\sigma(T)$ up to 950 K temperature of Cooper, silver and gold containing materials, correspondingly in constant relaxation time ($\tau$). According to the Fig. 9(d) and (e), the cooper and silver containing materials show the linearly increasing temperature dependent electrical conductivity, whereas in fig. 9(f), Au-based compounds exhibit almost temperature independent electrical conductivity. The linearly increasing trend of $\sigma(T)$ with temperature follows as $\sigma(CNT) > \sigma(CNSe) > \sigma(CNS)$ in Cu-based compounds whereas $\sigma(ANS) > \sigma(ANSe) > \sigma(ANT)$, and $\sigma(AuNT) > \sigma(AuNS) > \sigma(AuNSe)$ in Au-based compounds. The higher incident EMR energy generated additional electrons-holes in materials which leads overall the electrical conductivity $\sigma(T)$. The calculated electronic ($\kappa_e$) and lattice ($\kappa_L$) thermal conductivity with changing temperature up to 900 K are plotted in Fig. 10(a-f) of $X_3NbY_4$(X= Cu, Ag, Au; Y=S, Se, Te) compounds, respectively.

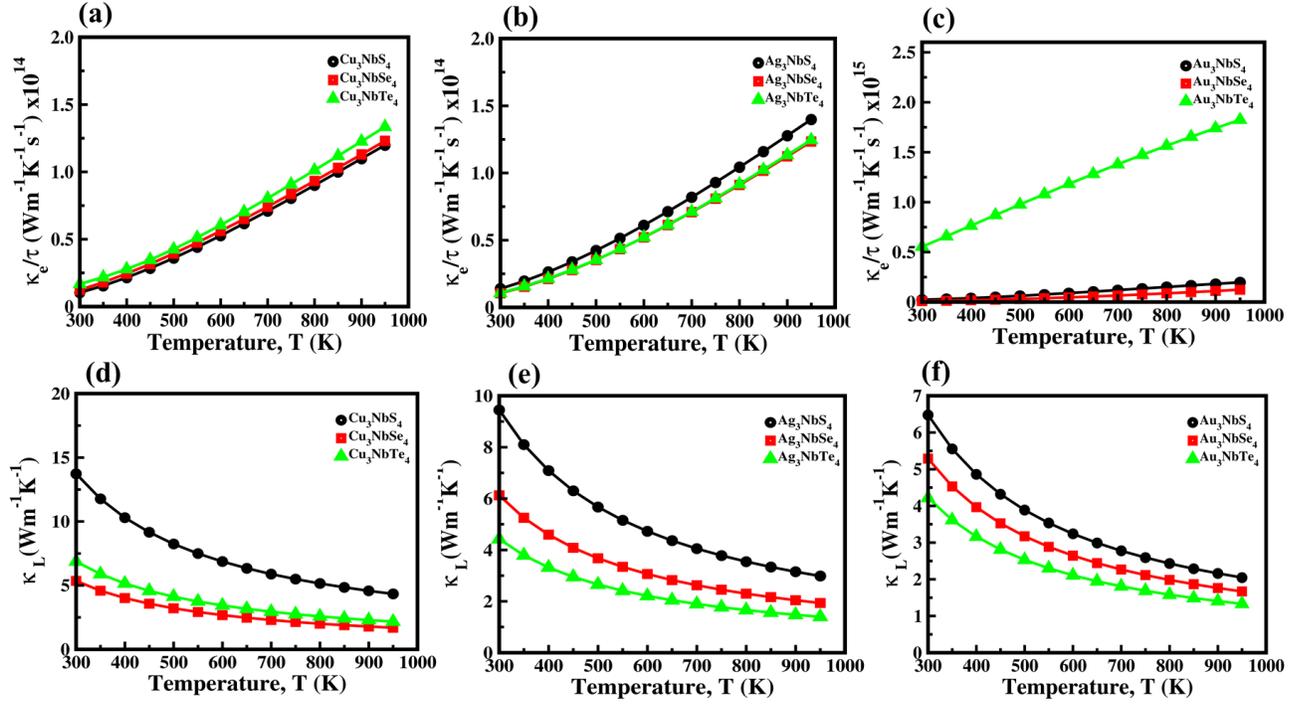

**Fig. 10:** Show the variation (a-c) of conductivity (electronic thermal) $\kappa_e(T)$ and (d-f) of lattice $\kappa_L(T)$ conductivity (thermal) up to 950 K temperature for the Cu-, Ag- and Au- based compounds, respectively, where CNS, ANS, AuNS in black curves, CNSe, ANSe, AuNSe in red curves, CNT, ANT, AuNT in green curves, respectively.

The value of $\kappa_e$ increases and $\kappa_L$ decreases with rising the temperature for all semiconducting compounds. Because in semiconducting materials, the electronic and thermal conductivity are significantly influence by the electronic motion as well as lattice oscillations. The total thermal conductivity is strongly affected by the electronic thermal conductivity but slightly affected by the lattice thermal conductivity. As shown in Fig. 10(a-f), the ratio of the value of $\kappa_e$ and $\kappa_L$ is estimated about $\sim 10^{14}$. As the Wiedemann–Franz law, the ratio of the electronic thermal conductivity to electrical conductivity is proportional to the temperature. Therefore, both of the conductivity (electronic thermal and electrical) is increases with temperature. With increasing temperature, the carrier concentrations in semiconductors materials increases owing to the thermal excitations because the electrons jump from valence band to conduction band. Hence, more carriers participate to transfer heat energy which leads the higher value of $\kappa_e$ about $\sim 10^{14}$ $Wm^1K^{-2}s^{-1}$.

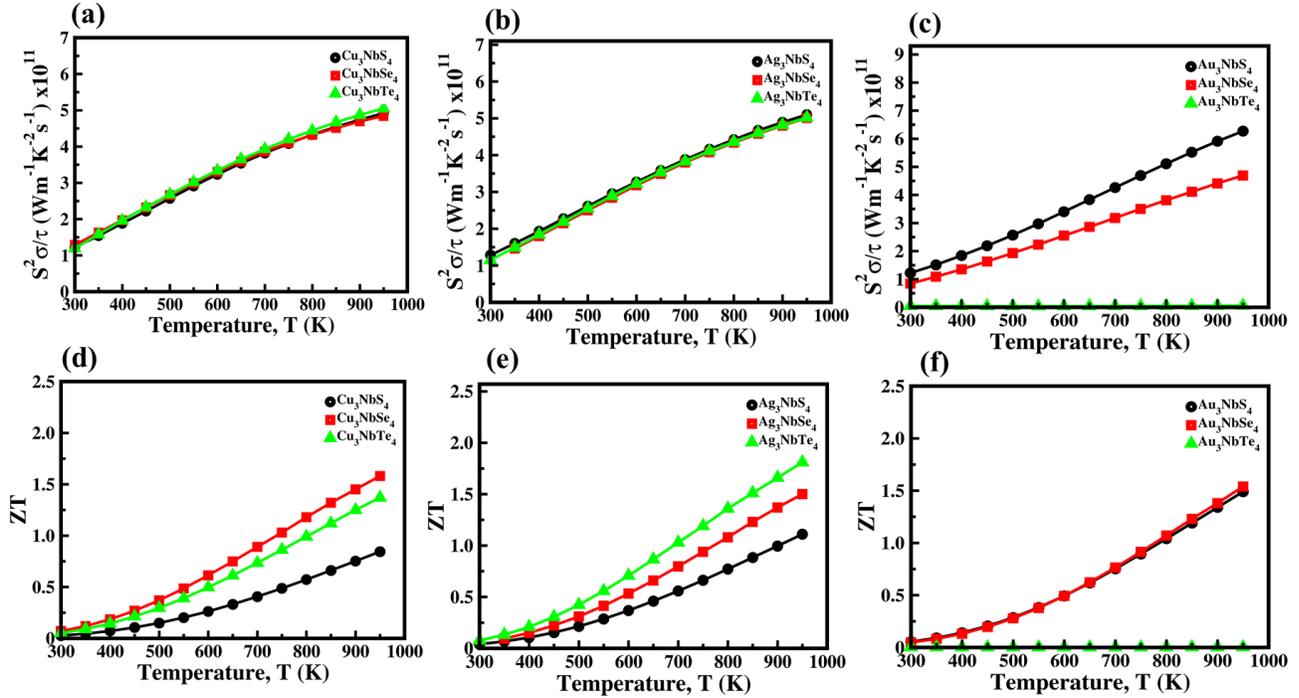

**Fig. 11:** Show the variation (a-c) of power factor P.F(T) and (d-f) of dimensionless of figure of merit ZT value with temperature up to 950 K for the Cu-, Ag- and Au- based compounds, respectively, Where CNS, ANS, AuNS in black curves, CNSe, ANSe, AuNSe in red curves, CNT, ANT, AuNT in green curves, respectively.

The calculated power factor (P.F.) up to 950 K temperature for all samples is plotted in the Fig. 11 (a-c) of the considered materials. According to the Fig. 11(a-c), the estimated value of P.F for all compounds except the Au$_3$NbTe$_4$ compound with temperature. A moderately higher value of P. F has been calculated ~ $10^{11}\ Wm^{-1}K^{-2}s^{-1}$ at higher temperatures in each compound. The recorded value of unitless quantity ZT value with temperature is presented in Fig. 11 (d-f) of Cu-, Ag- and Au-based compounds, respectively which measure the how much heat energy effectively converts into electrical energy. Fig. 11(d-f) display that at 300 K, the recorded value of ZT is very low, but its value increases with temperature for all compounds except Au$_3$NbTe$_4$ compound. The estimated ZT value at 600 K have been recorded about 0.31, 0.52, 0.50, 0.45, 0.62, 0.72, 0.50, and 0.50 for CNS, CNSe, CNT, ANS, ANSe, ANT, AuNS and AuNSe materials, correspondingly but the ZT vale for AuNT is negligible. According to the Fig. 11(d-f), highest ZT value at 950 K have been estimated approximately 0.82, 1.52, 1.45, 1.25, 1.52, 1.75, 1.50, and 1.52 for CNS, CNSe, CNT,

ANS, ANSe, ANT, AuNS and AuNSe compounds, respectively. The larger value of ZT of TEMs suggesting that these materials show the better TE efficiency with less heat energy dissipation. The value of ZT for all compounds significantly changes with the replacement of $Cu \rightarrow Ag \rightarrow Au$ atoms as well as S → Se → Te atoms. It is also observed that the ZT value follows increasing trend of electrical and electronic thermal conductivity. These results relatively higher value of Seebeck coefficient and electrical conductivity and effective value of ZT suggest that these semiconducting materials are potential TEMs for the higher temperature devices applications including the waste heat recovery technology.

## 4. Conclusion

The electronic behavior, elastic, optical, and transport properties of $X_3NbY_4$(X= Cu, Ag, Au; Y=S, Se, Te) chalcogenides materials using first-principles calculations based on the DFT have been investigated broadly via wien2k code. The estimated band structure nature and bandgap value in the range from 1.65 eV to 0.50 eV using PBE-GGA functional and 1.80 eV to 1.18 eV via TB-mBJ functional indicates that these compounds are indirect bandgap type semiconductor. These results highlight their importance as absorber materials in photovoltaic cells. The density of state ensures the structural stability and the investigation of PDOS value approves the valence band maximum mainly arises due to the X-*d*/Nb-*d*/Y-*p* hybridized of atomic orbitals whereas the conduction band maximum mostly attributes owing to hybridized of Nb-*d*/Y-*p* atomic orbitals. The estimated optical properties exhibit that the considered compounds are potential candidates in the energy harvesting applications such as optoelectronic devices and photovoltaic cells. The computed higher absorption coefficient (~$10^5$ cm$^{-1}$) of the materials indicates that these Cu-/Ag-/Au-based chalcogenides play the potential role for the energy harvesting technology. The estimated binding energy as well as the Bohr radius of Cu-/Ag-/Au-based compounds are the suitable for PV and optoelectronic devices industries whereas the Cu-/Ag-/Au containing materials are appropriate for the LEDs and solar cells deceives. Moreover, broadly investigation of Seebeck coefficient, lattice as well as electronic thermal conductivity, electrical conductivity, power factor (P.F) and dimensionless quantity figure of merit ZT value measure their importance in thermoelectric applications. Our finding through this research explores that theses chalcogenide materials potential for the technological applications and demands further experimental investigations as well as validations of $X_3NbY_4$(X= Cu, Ag, Au; Y=S, Se, Te) materials.

***Corresponding Author:*** E-mail: rashidphysics18@ru.ac.bd (Md. Abdur Rashid).

E-mail: jak_apee@ru.ac.bd (Jaker Hossain).

**Disclosure:** None of the authors have any competing financial interest.

**Data availability:** Data will be available from the corresponding author upon reasonable request.

**References**

[1] Zhang, Yubo, et al. Structural properties and quasiparticle band structures of Cu-based quaternary semiconductors for photovoltaic applications. *Journal of Applied Physics* 111.6 (2012). https://doi.org/10.1063/1.3696964

[2] Kehoe, Aoife B., David O. Scanlon, and Graeme W. Watson. Modelling potential photovoltaic absorbers $Cu_3MCh_4$ (M= V, Nb, Ta; Ch= S, Se, Te) using density functional theory. *Journal of Physics: Condensed Matter* 28.17 (2016): 175801. http://dx.doi.org/10.1088/0953-8984/28/17/175801

[3] Kehoe, Aoife B., David O. Scanlon, and Graeme W. Watson. The electronic structure of sulvanite structured semiconductors $Cu_3MCh_4$ (M= V, Nb, Ta; Ch= S, Se, Te): prospects for optoelectronic applications. *Journal of Materials Chemistry C* 3.47 (2015): 12236-12244. https://doi.org/10.1039/C5TC02760H

[4] Chen, Erica M., et al. Tuning the optical, electronic and thermal properties of $Cu_3NbS_{4-x}Se_x$ through chemical substitution. *Inorganic Chemistry Frontiers* 4.9 (2017): 1493-1500. https://doi.org/10.1039/C7QI00264E

[5] Rashid, Md Abdur, et al. Theoretical insights into the electronic structures and transport properties of Ag3TaX4 (X= S, Se, Te) compounds for energy applications. *Next Materials* 10

(2026): 101438. https://doi.org/10.1016/j.nxmate.2025.101438

[6] Aman, Md Sharear, et al. Electronic, optical and transport properties of ternary $\alpha_3V\beta_4$ ($\alpha$= Cu, Ag, and $\beta$= S, Se, Te) for energy harvesting: A DFT insights. Materials Today Communications, 51 (2026) 114873. https://doi.org/10.1016/j.mtcomm.2026.114873

[7] Rifat, Istiwak Ahammed, et al. Comprehensive Investigations on the Electronic and Transport Properties of $Au_3VX_4$ (X= S, Se, Te) Compounds via DFT Study for Applications in TPV Cells. *ACS Applied Engineering Materials* (2026). https://doi.org/10.1021/acsaenm.5c01044

[8] Green, Martin, et al. Solar cell efficiency tables (version 57). *Progress in photovoltaics: research and applications* 29.1 (2021): 3-15. https://doi.org/10.1002/pip.3371

[9] Shockley, William, and Hans Queisser. Detailed balance limit of efficiency of p–n junction solar cells. *Renewable energy*. Routledge, 2018. Vol2-35-Vol2_54.

[10] J. Tate, P. F. Newhouse, R. Kykyneshi, P. A. Hersh, J. Kinney, D. H. McIntyre, D. A. Keszler, Thin Solid Films 516, 5795 (2008). http://dx.doi.org/10.1016/j.tsf.2007.10.073

[11] P. F. Newhouse, P.A. Hersh, A. Zakutayev, A. Richard, H. A. S. Platt, D. A. Keszler, J. Tate, Thin Solid Films 517, 2473 (2009). http://dx.doi.org/10.1016/j.tsf.2008.11.020

[12] Pauling, L.; Hultgren, R. The Crystal Structure of Sulvanite, $Cu_3VS_4$. Z. Kristallogr. - Cryst. Mater. 1933, 84, 204−212. https://doi.org/10.1524/zkri.1933.84.1.204

[13] Arribart, H.; Sapoval, B.; Gorochov, O.; LeNagard, N. Fast ion transport at room temperature in the mixed conductor Cu3VS4. Solid State Commun. 1978, 26, 435−439. https://doi.org/10.1016/0038-1098(78)90522-7

[14] Altermatt, P. P.; Kiesewetter, T.; Ellmer, K.; Tributsch, H. Specifying targets of future research in photovoltaic devices containing pyrite ($FeS_2$) by numerical modelling. Sol. Energy Mater. Sol. Cells 2002, 71, 18. https://doi.org/10.1016/S0927-0248(01)00053-8
27

[15] J. Li, H.-Y. Guo, D.M. Proserpio, A. Sironi, Exploring Tellurides: Synthesis and Characterization of New Binary, Ternary, and Quaternary Compounds, J. Solid State Chem. 117 (1995) 247–255. https://doi.org/10.1006/jssc.1995.1270.

[16] M. Kars, A. Rebbah, H. Rebbah, $Cu_3NbS_4$, Acta Crystallogr. Sect. E Struct. Reports Online. 61 (2005) i180–i181. https://doi.org/10.1107/S1600536805022397.

[17] Y.-J. Lu, J.A. Ibers, Synthesis and Characterization of $Cu_3NbSe_4$ and $KCu_2TaSe_4$, J. Solid State Chem. 107 (1993) 58–62. https://doi.org/10.1006/jssc.1993.1323.

[18] G.E. Delgado, A.J. Mora, S. Durán, M. Muñoz, P. Grima-Gallardo, Structural characterization of the ternary compound $Cu_3TaSe_4$, J. Alloys Compd. 439 (2007) 346–349. https://doi.org/10.1016/j.jallcom.2006.08.232.

[19] W.F. Espinosa-García, C.M. Ruiz-Tobón, J.M. Osorio-Guillén, The elastic and bonding properties of the sulvanite compounds: A first-principles study by local and semi-local functionals, Phys. B Condens. Matter. 406 (2011) 3788–3793. https://doi.org/10.1016/j.physb.2011.06.060.

[20] G.E. Delgado, J.E. Contreras, A.J. Mora, S. Durán, M. Muñoz, P. Grima-Gallardo, Structure Refinement of the Semiconducting Compound $Cu_3TaS_4$ from X-Ray Powder Diffraction Data, Acta Phys. Pol. A. 120 (2011) 468–472. https://doi.org/10.12693/APhysPolA.120.468

[21] Liu, Xiao-Peng, et al. Promising thermoelectric materials of $Cu_3VX_4$ (X= S, Se, Te): A Cu-VX framework plus void tunnels. *International Journal of Modern Physics C* 30.08 (2019): 1950045. https://doi.org/10.1142/S0129183119500451

[22] Espinosa-García, W. F., et al. The electronic and optical properties of the sulvanite compounds: a many-body perturbation and time-dependent density functional theory study. *Journal of Physics: Condensed Matter* 30.3 (2017): 035502. https://DOI 10.1088/1361-648X/aa9deb

[23] J.M. Osorio-Guillén, W.F. Espinosa-García, A first-principles study of the electronic structure of the sulvanite compounds, Phys. B Condens. Matter. 407 (2012) 985–991. https://doi.org/10.1016/j.physb.2011.12.126.

[24] Arribart, H., and Bernard Sapoval. Theory of mixed conduction due to cationic interstitials in the p-type semiconductor $Cu_3VS_4$. *Electrochimica Acta* 24.7 (1979): 751-754. https://doi.org/10.1016/0013-4686(79)80004-3

[25] Petritis, D., et al. Investigation of the vibronic properties of $Cu_3VS_4$, $Cu_3NbS_4$, and $Cu_3TaS_4$ compounds. *Physical Review B* 23.12 (1981): 6773. https://doi.org/10.1103/PhysRevB.23.6773.

[26] M.C. Payne, M.P. Teter, D.C. Allan, T.A. Arias, J.D. Joannopoulos, Iterative minimization techniques for ab initio total-energy calculations: molecular dynamics and conjugate gradients. https://doi.org/10.1103/RevModPhys.64.1045

[27] Tran, Fabien, Peter Blaha, and Karlheinz Schwarz. Band gap calculations with Becke–Johnson exchange potential. *Journal of Physics: Condensed Matter* 19.19 (2007): 196208. DOI 10.1088/0953-8984/19/19/196208

[28] Tran, Fabien, and Peter Blaha. Accurate band gaps of semiconductors and insulators with a semilocal exchange-correlation potential. *Physical review letters* 102.22 (2009): 226401. https://doi.org/10.1103/PhysRevLett.102.226401

[29] G.K.H. Madsen, D.J. Singh, BoltzTraP. A code for calculating band-structure dependent quantities, Comput. Phys. Commun. 175 (2006) 67–71. https://doi.org/10.1016/j.cpc.2006.03.007.

[30] J.P. Perdew, K. Burke, M. Ernzerhof, Generalized Gradient Approximation Made Simple, Phys. Rev. Lett. 77 (1996) 3865–3868. https://doi.org/10.1103/PhysRevLett.77.3865.

[31] Perdew, John P., et al. Restoring the density-gradient expansion for exchange in solids and surfaces. Physical *review letters* 100.13 (2008): 136406. https://doi.org/10.1103/PhysRevLett.100.136406.

[32] H.J. Monkhorst, J.D. Pack, Special points for Brillouin-zone integrations, Phys. Rev. B. 13 (1976) 5188–5192. https://doi.org/10.1103/PhysRevB.13.5188.

[33] Ikeda, Yuji, et al. Mode decomposition based on crystallographic symmetry in the band-unfolding method. *Physical Review B* 95.2 (2017): 024305. https://doi.org/10.1103/PhysRevB.95.024305

[34] C. Mujica, G. Carvajal, J. Llanos, O. Wittke, Redetermination of the crystal structure of copper(I) tetrathiovanadate (sulvanite), $Cu_3VS_4$, Zeitschrift Für Krist. - New Cryst. Struct. 213 (1998) 12. https://doi.org/10.1524/ncrs.1998.213.14.12

[35] Matyszczak, Grzegorz, et al. Synthesis, characterization, crystal structure prediction, and ab initio study of bandgap of Cu3VSe4. *Journal of Solid State Chemistry* 301 (2021): 122336. https://doi.org/10.1016/j.jssc.2021.122336.

[36] Wen, Jiahao, et al. Thermoelectric properties of p-Type $Cu_3VSe_4$ with high Seebeck coefficients. *Journal of Alloys and Compounds* 879 (2021): 160387. https://doi.org/10.1016/j.jallcom.2021.160387

[37] Alkhaldi, Hanof Dawas. Investigation of optical and thermoelectric characteristics of novel zintl-phase alloys $CaZn_2X_2$ (X= P, As, Sb) for green energy applications. *Physica Scripta 99.12* (2024): 125991. DOI 10.1088/1402-4896/ad935f

[38] Hong, A. J., et al. Novel p-type thermoelectric materials $Cu_3MCh_4$ (M= V, Nb, Ta; Ch= Se, Te): high band-degeneracy. *Journal of Materials Chemistry A* 5.20 (2017): 9785-9792. https://doi.org/10.1039/C7TA02178J

[39] Mohamed, Abdelhay Salah, et al. Investigating novel Mo$_2$X$_3$S (X= Se, Te) materials: probing the influence of chalcogen substitution on electronic, optical, and thermoelectric properties. *Physica Scripta* 99.10 (2024): 105949. DOI 10.1088/1402-4896/ad7420

[40] K. Bougherara, F. Litimein, R. Khenata, E. Uçgun, H.Y. Ocak, Ş. Uğur, G. Uğur, A. Reshak, F. Soyalp, S. Bin Omran, Structural, Elastic, Electronic and Optical Properties of Cu$_3$TMSe$_4$ (TM = V, Nb and Ta) Sulvanite Compounds via First-Principles Calculations, Sci. Adv. Mater. 5 (2013) 97–106. https://doi.org/10.1166/sam.2013.1435.

[41] Mouhat, Félix, and François-Xavier Coudert. Necessary and sufficient elastic stability conditions in various crystal systems. *Physical review B* 90.22 (2014): 224104. https://doi.org/10.1103/PhysRevB.90.224104.

[42] W. Feng, S. Cui, Mechanical and electronic properties of Ti$_2$AlN and Ti$_4$AlN$_3$ : a first-principles study, Can. J. Phys. 92 (2014) 1652–1657. https://doi.org/10.1139/cjp-2013-0746.

[43] D.G. Pettifor, Theoretical predictions of structure and related properties of intermetallics, Mater. Sci. Technol. 8 (1992) 345–349. https://doi.org/10.1179/mst.1992.8.4.345.

[44] S.F. Pugh, XCII. Relations between the elastic moduli and the plastic properties of polycrystalline pure metals, London, Edinburgh, Dublin Philos. Mag. J. Sci. 45 (1954) 823–843, https://doi.org/10.1080/14786440808520496.

[45] G.N. Greaves, A.L. Greer, R.S. Lakes, T. Rouxel, Poisson's ratio and modern materials, Nat. Mater. 10 (2011) 823–837. https://doi.org/10.1038/nmat3134.

[46] Ahmadi, Aidin, et al. Thermodynamic, Mechanical, Optical and Electronic Properties of Cu$_3$VS$_4$: An Ab Initio Study: Ahmadi, Nouri, Taghizade, and Faghihnasiri. *Journal of Electronic Materials* 50.1 (2021): 336-345. https://doi.org/10.1007/s11664-020-08557-1

[47] Voigt, W. Lehrbuch der kristallphysik teubner, leipzig (1928).

[48] Reuss, A., and Z. Angnew. A calculation of the bulk modulus of polycrystalline materials. *Math Meth* 9 (1929): 55.

[49] Hill, R. Proc. Phys. Soc., London. *Sect. A* 65 (1952): 349.

[50] Alkhaldi, Hanof Dawas. Investigation of optical and thermoelectric characteristics of novel zintl-phase alloys $CaZn_2X_2$ (X= P, As, Sb) for green energy applications. *Physica Scripta 99.12* (2024): 125991. DOI 10.1088/1402-4896/ad935f

[51] Hong, A. J., et al. Novel p-type thermoelectric materials $Cu_3MCh_4$ (M= V, Nb, Ta; Ch= Se, Te): high band-degeneracy. *Journal of Materials Chemistry A* 5.20 (2017): 9785-9792. https://doi.org/10.1039/C7TA02178J

[52] Mohamed, Abdelhay Salah, et al. Investigating novel $Mo_2X_3S$ (X= Se, Te) materials: probing the influence of chalcogen substitution on electronic, optical, and thermoelectric properties. *Physica Scripta* 99.10 (2024): 105949. DOI 10.1088/1402-4896/ad7420

[53] Rani, Upasana, et al. Fundamental theoretical design of Na- ion and K- ion based double antiperovskite $X_6SOA_2$ (X= Na, K; A= Cl, Br and I) halides: potential candidate for energy storage and harvester. *International Journal of Energy Research* 45.9 (2021): 13442-13460. https://doi.org/10.1002/er.6673

[54] Ahmadi, Aidin, et al. Thermodynamic, Mechanical, Optical and Electronic Properties of $Cu_3VS_4$: An Ab Initio Study: Ahmadi, Nouri, Taghizade, and Faghihnasiri. *Journal of Electronic Materials* 50.1 (2021): 336-345. https://doi.org/10.1007/s11664-020-08557-1

[55] Mark Fox, optical properties of solids, Second edi, Clarendon Press, Oxford, 2010.

[56] H. Joshi, D.P. Rai, L. Hnamte, A. Laref, R.K. Thapa, A theoretical analysis of elastic and optical properties of half Heusler MCoSb (M=Ti, Zr and Hf), Heliyon 5 (2019) e01155. https://doi.org/10.1016/j.heliyon.2019.e01155.

[57] Lv, X. S., et al. Fundamental optical and electrical properties of nano-$Cu_3VS_4$ thin film. *Optical Materials* 34.8 (2012): 1451-1454. https://doi.org/10.1016/j.optmat.2012.02.044

[58] Saddique, Jaffer, et al. First-principles investigation of structural, electronic, optical, and mechanical properties of novel metallic ternary halides $K_3XBr_6$ (X= Sc, Y). *Results in Physics* (2025): 108347. https://doi.org/10.1016/j.rinp.2025.108347

[59] Jang, Hyun Myung, et al. Exciton Bohr radius of lead halide perovskites for photovoltaic and light-emitting applications. *arXiv preprint arXiv:2505.15565* (2025). https://doi.org/10.48550/arXiv.2505.15565

[60] Dvorak, M., Wei, S. H., & Wu, Z. (2013). Origin of the variation of exciton binding energy in semiconductors. *Physical review letters*, *110*(1), 016402. https://doi.org/10.1103/PhysRevLett.110.016402

[61] Lee, J. C., Chai, J. D., & Lin, S. T. (2015). Assessment of density functional methods for exciton binding energies and related optoelectronic properties. *RSC Advances*, *5*(123), 101370-101376. https://doi.org/10.1039/C5RA20085G

[62] Qin, Guangzhao, et al. High-throughput computational evaluation of lattice thermal conductivity using an optimized Slack model. *Materials Advances* 3.17 (2022): 6826-6830. https://doi.org/10.1039/D2MA00694D

[63] Allen, Philip B., Warren E. Pickett, and Henry Krakauer. Anisotropic normal-state transport properties predicted and analyzed for high-T c oxide superconductors. *Physical Review B* 37.13 (1988): 7482. https://doi.org/10.1103/PhysRevB.37.7482

[64] Scheidemantel, T. J., et al. Transport coefficients from first-principles calculations. *Physical Review B* 68.12 (2003): 125210. https://doi.org/10.1103/PhysRevB.68.125210

[65] Khan, Muhammad Salman, et al. A computational study of novel CsYMSe$_3$ (M= Cd, Zn) materials: for their potential optoelectronic and thermoelectric application. *Modelling and Simulation in Materials Science and Engineering* 33.2 (2025): 025005. DOI 10.1088/1361-651X/ada81d